\documentclass[lettersize,journal]{IEEEtran}
\usepackage{amsmath,amsfonts}
\usepackage{algorithm}
\usepackage{algpseudocode}
\usepackage{array}
\usepackage{multirow}
\usepackage{textcomp}
\usepackage{stfloats}
\usepackage{url}
\usepackage{verbatim}
\usepackage{graphicx}
\usepackage{cite}
\usepackage{caption} 
\usepackage{amsmath}
\usepackage{xspace}
\usepackage{fixltx2e}
\usepackage{subcaption}

\graphicspath{{Figure/}}

\begin{document}

\title{Secrecy Capacity of Hybrid VLC-RF Systems with Light Energy Harvesting}

\author{Tuan A. Hoang, Thanh V. Pham, ~\IEEEmembership{Senior Member,~IEEE}, Chuyen T. Nguyen

\thanks{Tuan A. Hoang and Chuyen T. Nguyen are with the School of Electrical and Electronic Engineering, Hanoi University of Science and Technology, Hanoi, Vietnam (email: tuan.ha203826@sis.hust.edu.vn, chuyen.nguyenthanh@hust.edu.vn).

Thanh V. Pham with the Department of Mathematical and Systems Engineering, Shizuoka University, Shizuoka, Japan (e-mail: pham.van.thanh@shizuoka.ac.jp). 
}}
\maketitle
\begin{abstract}
This paper studies the performance of physical layer security (PLS) in a multi-user hybrid heterogeneous visible light communication (VLC) and radio frequency (RF) wireless communication system with simultaneous lightwave information and power transfer (SLIPT). In the considered system, VLC is used for downlink (DL) while RF is employed for uplink (UL) transmission. In addition, to support multiple users, time division multiple access (TDMA) is assumed for both DL and UL channels. In the DL, each user receives information during its allocated time slot of the TDMA frame and harvests energy from the received signal outside the time slot. The harvested energy is then used for transmitting the signal over the UL channel, which is subject to eavesdropping by an unauthorized user. Therefore, PLS is employed to protect the confidentiality of the UL information. Then, an optimization problem is formulated to solve the optimal DL and UL time slots that maximize the PLS performance given a target sum rate of the DL. We show that the problem can be cast as a difference of convex functions (DC) program, which can be solved efficiently using the DC algorithm (DCA).  
\end{abstract}


\section{Introduction}
Due to the explosion of data traffic and the congestion of the radio frequency (RF) spectrum, visible light communication (VLC) is emerging as a promising technology for the future of wireless communications \cite{ariyanti2020visible}.  VLC offers the potential for significantly higher achievable data rates, largely thanks to its abundance of unlicensed spectrum \cite{tavakkolnia2018energy} and its ability to avoid interference with existing radio frequency (RF) systems \cite{miranda2023review}. 
However, despite these advantages, VLC has a significant drawback compared to traditional RF systems. The coverage area of LED illumination, which VLC relies on, is limited and can be easily obstructed by physical objects \cite{zjovicic2013visible}. This dependency on line-of-sight transmission means that VLC signals can be easily blocked by walls, furniture, or even people, resulting in potential communication blackouts or interruptions. VLC is also impractical for uplink (UL) communications because it is not feasible for phones or computers to emit light to transmit data.
To address this issue, researchers have proposed hybrid VLC-RF systems \cite{7848882} where the DL utilizes a VLC channel to take advantage of its high data rate capabilities and the UL employs a traditional RF channel. 

In such hybrid VLC-RF systems, while the DL VLC channel is considered secure due to the confinement of the visible light (typically within a room), the UL RF transmission can be easily wiretapped by malicious users due to the broadcast nature of the radio signals. Traditionally, encryption methods implemented at the upper layers of communication protocols (i.e., network layer and above) have been employed to secure data transmission. These methods leverage advanced cryptographic theory to encode data such that with the current computing capability, an unauthorized user cannot decode the encrypted message within a reasonable amount of time without the secret key.
However, advances in the development of quantum computers mean that traditional cryptographic algorithms can be compromised \cite{Castelvecchi2022}. Hence, information-theoretic security, such as 
physical layer security (PLS), can be seen as a promising complementary approach to deal with attackers having unlimited computing power \cite{Wyner1975}. The performance of PLS can be numerically quantified by the so-called secrecy capacity, which is the maximum transmission rate at which information can be securely transmitted from a transmitter (Alice) to a legitimate user (Bob) with the presence of an eavesdropper (Eve). 

In recent years, there has been an increasing interest in applying hybrid VLC-RF systems for the Internet of Things (IoT) and wireless sensor networks (WSN) which often comprise a large number of devices with limited battery power \cite{Delgado2020}. As a result, frequent recharging or battery replacement is required, which can be both inconvenient and costly, particularly in large-scale deployments. To address this issue, energy harvesting (EH) can be considered an effective technique to 
prolong battery life by converting the received signal into electrical power, thus reducing the dependency on conventional power sources. In the considered hybrid VLC-RF systems, user devices harvest the energy from the received VLC signal by means of simultaneous lightwave information and power transfer (SLIPT) \cite{papanikolaou2023simultaneous} and use the harvested energy for the UL transmission. Specifically, under the assumption that time division multiple access (TDMA) is employed for both UL and DL channels, each user device receives information during its allocated DL time slot and harvests energy outside that. To maximize the secrecy capacity of the UL transmission given a predefined target sum rate of the DL, an optimization problem is formulated to solve the optimal DL and UL time slots for all user devices. The problem is shown to be a difference of convex functions (DC) program, which can be efficiently solved using the DC algorithm (DCA) \cite{tao1997convex}.
\section{System Model}
\begin{figure}[ht]
    \centering
    \includegraphics[width=0.45\textwidth, height = 5cm]{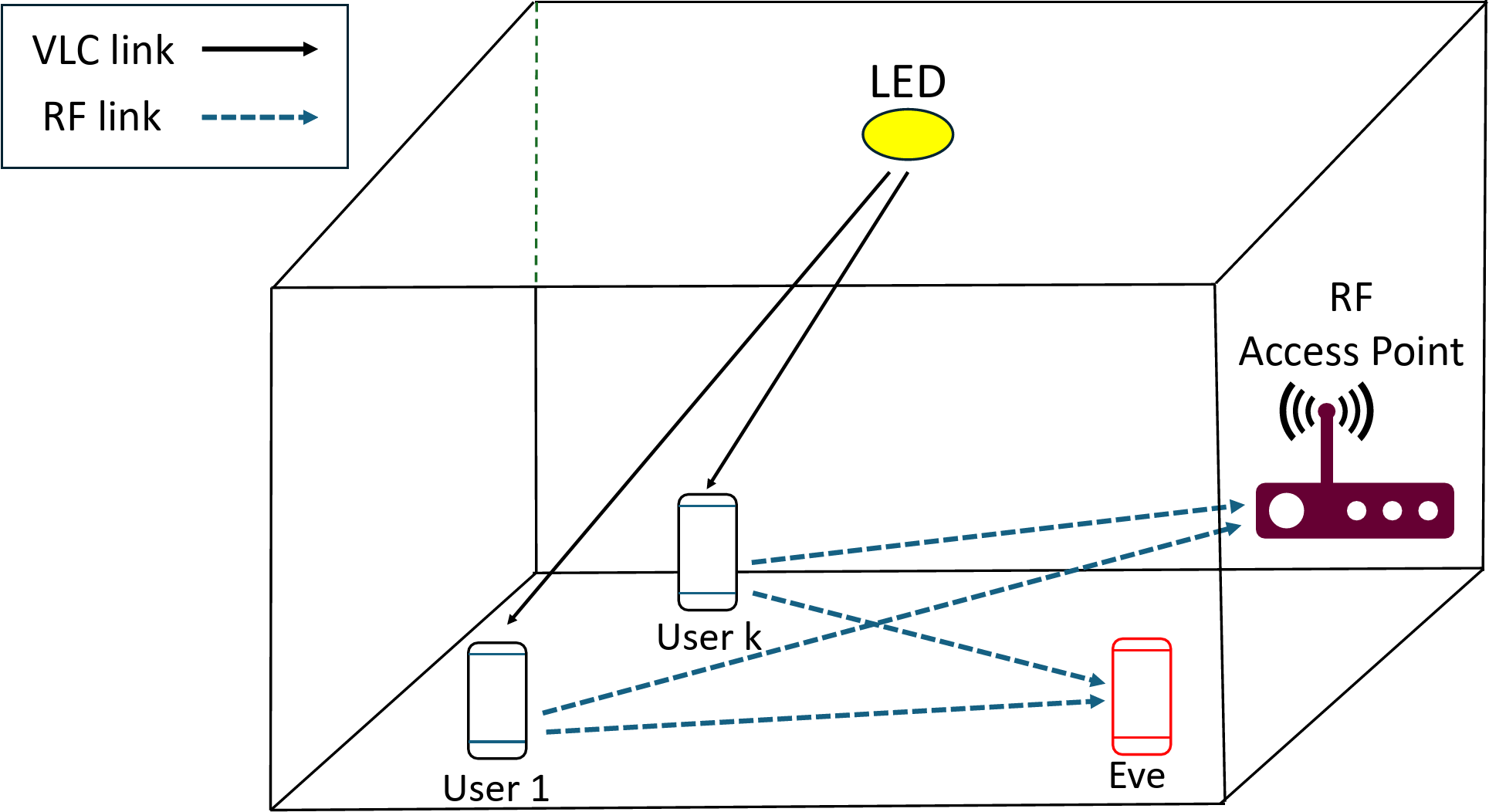}
    \caption{Hybrid VLC-RF system model.}
    \label{fig:1}
\end{figure} 
\noindent
The considered hybrid VLC-RF system, as depicted in Fig.~\ref{fig:1}, features a single LED transmitter and an RF access point, both capable of serving $K$ ($K > 1$) users. The LED transmitter is centrally positioned on the ceiling, while the users are randomly distributed throughout the room. Furthermore, in the downlink, the LED transmitter provides energy for users by using Simultaneous Lightwave Information and Power Transfer (SLIPT). For the user $k$ (k = 1, 2, ..., $K$), the optical signal transmitted by the LED can be expressed as
\begin{equation}
x_{k} = \sqrt{P_{\text{LED}}} s_{k} + I_{\text{D}},
\label{eq:1}
\end{equation}
where $P_{\text{LED}}$ is the transmitted power of the LED, $s_{k}$ is the transmitted data of user $k$, and $I_{\text{D}}$ is the DC offset which controls the brightness of the LED. Therefore, the electrical signal received by the user k is written by 
\begin{equation}
y_{k} = \xi g_{k} x_{k} + n_{k},
\label{eq:2}
\end{equation}
where $\xi$ is the optical-electrical conversion efficiency, $n_{k}$ is additive white Gaussian noise (AWGN) which primarily originates from shot noise and thermal noise, and $g_{k}$ is the channel gain of the VLC link, which can be represented as \cite{komine2004fundamental}
\begin{equation}
g_{k} = \frac {(m+1) A_{P} R_{P}}{2 \pi d_{k}^2} \cos^m(\phi_{k}) T_{s}(\varphi_{k}) T_{f}(\varphi_{k}) \cos (\varphi_{k}),
\label{eq:3}
\end{equation}
with $0\leq\phi_{k}\leq\Psi$ and $g_{k} = 0$ otherwise. In \eqref{eq:3}, $m = -1/\log_{2}(\cos(\phi_{1/2}))$, with $\phi_{1/2}$ being the semi-angle at half illuminance of the LED, is the Lambertian radiation, $\phi_{k}$ and $\varphi_{k}$ denote the angles of irradiance and incidence between the LED and each user, $\Psi$ denotes the photodiode’s field of view (FOV), $d_{k}$ is the distance between the LED and user $k$, $A_{P}$ and $R_{P}$ represent the detection area and responsivity of the PD, respectively. Also, $T_{s}(\varphi_{k})$ represents the gain of the optical filter, and $T_{f}(\varphi_{k})$ represents the non-imaging concentrator gain, which is given by:
\begin{equation}
T_{f}(\varphi_{k}) = \begin{cases}
  \frac {\kappa^2}{\sin^2(\varphi_{k})}, &\quad 0 \le \varphi_{k} \le \Psi ,\\
  0, &\quad \text{otherwise},
\end{cases} 
\label{eq:4}
\end{equation}
where $\kappa$ is the refractive index.

\begin{figure}[ht]
    \centering
    \includegraphics[width=0.45\textwidth]{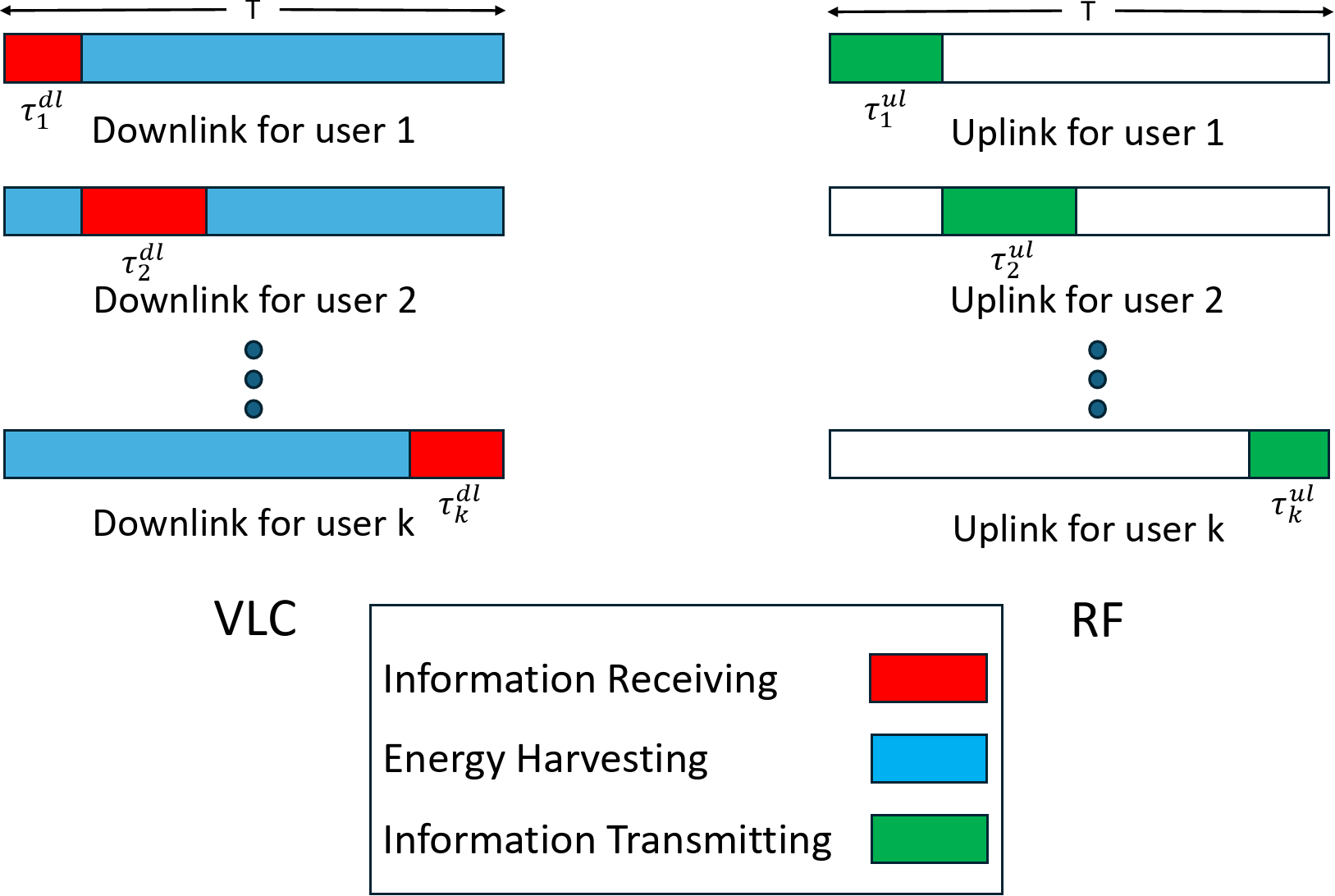}
    \caption{Time frame for $K$ users in hybrid VLC-RF system \cite{zargari2021resource}.}
    \label{fig:2}
\end{figure}

To support multiple users, the considered hybrid VLC-RF system employs TDMA for both DL and UL transmissions. Specifically, as shown in  Fig.~\ref{fig:2}, a TDMA frame of $T$ seconds is divided into $K$ time slots where $\tau_k^{\text{dl}}$ and $\tau_k^{\text{ul}}$  are the DL and UL time slots of the user $k$ \cite{zargari2021resource}. In the DL, the user $k$ receives information during the time slot $\tau_k^{\text{dl}}$ and harvests energy from the received VLC signal during the time slot $1 - \tau_k^{\text{dl}}$. 
\noindent
Note that each user can simultaneously receive and transmit signals over both VLC and RF links. 
For simplicity, we set $T = 1$. The energy harvested from the VLC link at user $k$ is
\begin{equation}
E_{k} = \eta I_{\text{D}}^2 g_{k}^2(1-\tau_{k}^{\text{dl}}),
\label{eq:5}
\end{equation}
with $\eta$ representing the energy conversion efficiency. This harvested energy is utilized for transmitting signals in the UL in the portion of time $\tau^{\text{ul}}_k$. The power used for the UL transmission is thus given by
\begin{equation}
P_{k} = \frac {E_{k}} {\tau_{k}^{\text{ul}}} = \frac {\eta I_{\text{D}}^2 g_{k}^2(1-\tau_{k}^{\text{dl}})} {\tau_{k}^{\text{ul}}}. 
\label{eq:6}
\end{equation}

\section{UL and DL Time Slots Optimization}

This section focuses on maximizing the sum of secrecy capacity in the UL channel while attaining a targeted sum rate in the DL channel. An achievable sum rate of the DL channel can be calculated by \cite{wang2013tight}
\begin{equation}
R^{\text{dl}}_{\text{sum}} =  \sum_{k=1}^K \tau^{\text{dl}}_{k} \log_2\left( 1 + \frac {e} {2\pi} \frac {P_{\text{LED}} g^2_{k}} {\sigma_{k, \text{dl}}^2}\right), \label{eq:7}
\end{equation}
where $\sigma_{k, \text{dl}}^2$ is the noise power of the DL channel. For the UL, the channel secrecy capacity of user $k$ is expressed as the difference in capacities of the user's and the eavesdropper Eve's channels. Given the harvested power in \eqref{eq:6}, it is given by
\begin{align}
C_{S,k} = &  \tau_k^{\text{ul}} \log_2 \left( 1 + \frac{\eta I_{\text{D}}^2 g_{k}^2 (1-\tau_{k}^{\text{dl}}) h_{k}^2}{\tau_{k}^{\text{ul}} \sigma_{k, \text{dl}}^2} \right) \nonumber \\
& -  \tau_k^{\text{ul}} \log_2 \left( 1 + \frac{\eta I_{\text{D}}^2 g_{k}^2 (1-\tau_{k}^{\text{dl}}) h_{\text{E},k}^2}{\tau_{k}^{\text{ul}} \sigma_{\text{E}, \text{dl}}^2} \right),
\label{eq:12}
\end{align}
where $h_{k}$ ($h_{\text{E}}$) and $\sigma^2_{k, \text{dl}}$ ($\sigma^2_{\text{E}, \text{dl}}$) are the gain and noise power of the UL channel at the user $k$ (the eavesdropper Eve), respectively. Assuming the Rician distribution, $h_{k}$ ($h_{\text{E}}$) can be written by
\begin{equation}
h_k = \sqrt{\frac {K_r} {1+K_r}} h^{\text{LOS}}_k + \sqrt{\frac {K_r} {1+K_r}} h^{\text{NLOS}}_k,
\end{equation}
where $K_r$ is the Rician factor, $h^{\text{NLOS}}_k$ indicates the non-line-of-sight (NLoS) component which follows the Rayleigh fading model, and $h^{\text{LOS}}_k$ is the LoS component of user $k$.

Using \eqref{eq:7} and \eqref{eq:12}, our optimization problem can be formulated as follows
\begin{subequations}
\label{OptProb1}
    \begin{alignat}{2}
        &\underset {\tau_{k}^{\text{dl}} ; \tau_{k}^{\text{ul}}} {\text{maximize}} & & \sum_{k=1}^K C_{S,k} \hspace{2.2cm} \label{eq:8}\\
        &\text{subject to }  &&  \sum_{k=1}^K \tau_{k}^{\text{dl}} \leq 1, \hspace{1cm} \sum_{k=1}^K \tau_{k}^{\text{ul}} \leq 1, \label{eq:9} \\
        & & & \tau_{k}^{\text{dl}} \geq 0 , \hspace{0.5cm} \tau_{k}^{\text{ul}} \geq 0,\hspace{0.5cm} \forall k, \label{eq:10} \\
        & & & R^{\text{dl}}_{\text{sum}} \geq r_{\text{min}} \label{eq:11}.
    \end{alignat}
\end{subequations}
Equation (\ref{eq:9}) enforces the total time slot constraints for the DL and UL. Constraints in (\ref{eq:10}) indicate the non-negativity of the UL and DL time allocations. Equation (\ref{eq:11}) represents the sum rate in the downlink, where $r_{\text{min}}$ denotes the minimum sum rate. For conciseness, we denote that $u_k(\tau_k^{\text{dl}},\tau_k^{\text{ul}})=\tau_k^{\text{ul}} \log_2 \left( 1 + \frac{\eta I_{\text{D}}^2 g_{k}^2 (1-\tau_{k}^{\text{dl}}) h_{k}^2}{\tau_{k}^{\text{ul}} \sigma_{k, \text{dl}}^2} \right)$, $v_k(\tau_k^{\text{dl}},\tau_k^{\text{ul}})= \tau_k^{\text{ul}} \log_2 \left( 1 + \frac{\eta I_{\text{D}}^2 g_{k}^2 (1-\tau_{k}^{\text{dl}}) h_{\text{E},k}^2}{\tau_{k}^{\text{ul}} \sigma_{\text{E}, \text{dl}}^2}\right)$ and $a_k=\frac{\eta I_{\text{D}}^2 g_{k}^2 h_{k}^2}{\sigma_{k, \text{dl}}^2}$. We now prove that $u_k(\tau_k^{\text{dl}},\tau_k^{\text{ul}})$ and $v_k(\tau_k^{\text{dl}},\tau_k^{\text{ul}})$ are concave functions. For $u_k(\tau_k^{\text{dl}},\tau_k^{\text{ul}})$, its Hessian is given by 
\begin{equation}
\mathbf{H}_k = \begin{bmatrix}
\frac{\partial^2 u_k}{\partial (\tau_{k}^{\text{ul}})^2} & \frac{\partial^2 u_k}{\partial \tau_{k}^{\text{ul}} \partial \tau_{k}^{\text{dl}}} \\
\frac{\partial^2 u_k}{\partial \tau_{k}^{\text{ul}} \partial \tau_{k}^{\text{dl}}} & \frac{\partial^2 u_k}{\partial (\tau_{k}^{\text{dl}})^2}
\end{bmatrix} ,
\end{equation}
where
\begin{equation}
\frac{\partial^2 u_k}{\partial (\tau_{k}^{\text{dl}})^2} = \frac{-a_k^2\tau_k^{\text{ul}}}{(\tau_k^{\text{ul}} +  (1-\tau_{k}^{\text{dl}}) a_{k})^2 \log(2)}, 
\label{eq:14}
\end{equation}

\begin{equation}
\begin{aligned}
\frac{\partial^2 u_k}{\partial (\tau_{k}^{\text{ul}})^2} = & \frac{(1 - \tau_{k}^{\text{dl}}) a_{k}}{(\tau_{k}^{\text{ul}} + (1 - \tau_{k}^{\text{dl}}) a_{k})^2 \log(2)} \\
& + \frac{-(1 - \tau_{k}^{\text{dl}}) a_{k}}{\tau_{k}^{\text{ul}} (\tau_{k}^{\text{ul}} + (1 - \tau_{k}^{\text{dl}}) a_{k}) \log(2)},
\end{aligned}
\label{eq:15}
\end{equation}
and
\begin{equation}
\frac{\partial^2 u_k}{\partial \tau_{k}^{\text{ul}} \partial \tau_{k}^{\text{dl}}} = \frac{ -(1-\tau_{k}^{\text{dl}})a_k^2} {(\tau_{k}^{\text{ul}} + (1-\tau_{k}^{\text{dl}})a_{k})^2 \log(2)}.
\label{eq:16}
\end{equation}
For any real column vector $z = \begin{bmatrix} c_1 \\ c_2\end{bmatrix}$, we have

\begin{align}
&z^T\mathbf{H}_k z  = \frac{-1}{(\tau_k^{\text{ul}} +  (1-\tau_{k}^{\text{dl}}) a_{k})^2 \log(2)} \times \nonumber \\
&\left(a_k^2\tau_k^{\text{ul}}c_1^2 + (1 - \tau_{k}^{\text{dl}})a_k^22c_1c_2 +  \frac {((1-\tau_k^{\text{dl}})a_k)^2} {\tau_k^{\text{ul}}} c_2^2\right)
 \nonumber \\ 
& = \frac {-\left(a_k\sqrt{\tau_k^{\text{ul}}}c_1-\frac {\left(1-\tau_k^{\text{dl}}\right)a_k)} {\sqrt{\tau_k^{\text{ul}}}}\right)^2} {(\tau_{k}^{\text{ul}} + (1-\tau_{k}^{\text{dl}})a_{k})^2 \log(2)} \leq 0,
\end{align}
which shows that  $u_k(\tau_k^{\text{dl}},\tau_k^{\text{ul}})$ is a concave function. Similarly,  $v_k(\tau_k^{\text{dl}},\tau_k^{\text{ul}})$ is also a concave function. It can be observed that the objective function \eqref{eq:8} $\sum_{k=1}^K C_{S,k} = \sum_{k=1}^K u_k(\tau_k^{\text{dl}},\tau_k^{\text{ul}}) - \sum_{k=1}^K v_k(\tau_k^{\text{dl}},\tau_k^{\text{ul}})$ is a difference of two concave functions. The optimization problem in \eqref{OptProb1}, therefore, is a DC program, which can be efficiently solved using the DCA \cite{tao1997convex}. 
A pseudo-code for the DCA to solve \eqref{OptProb1} is shown in the $\textbf{Algorithm 1}$ as below.
\begin{algorithm}[ht]
\caption{Difference of Convex Algorithm (DCA)}
\begin{algorithmic}[1]
\State Let $x = \{\tau_1^{\text{dl}},\tau_1^{\text{ul}},...\tau_K^{\text{dl}},\tau_K^{\text{ul}}\}$,  $u(x) = \sum_{k=1}^K u_k(\tau_k^{\text{dl}}, \tau_k^{\text{ul}})$, $v(x) = \sum_{k=1}^K v_k(\tau_k^{\text{dl}}, \tau_k^{\text{ul}})$.
\State Initialize a starting point $x_0$ and an error tolerance $\epsilon$ for the algorithm 
\While{True}
     \State Find an optimal set $y_n \in \partial v(x_n)$
     \State Find an optimal set $x_{n+1} \in \partial u^*(y_n)$
     \If{$|x_{n+1}-x_n| \leq \epsilon$}
        \State return $x_{n+1}$
        \State break
    \EndIf
\EndWhile
\end{algorithmic}
\end{algorithm}

\noindent
In the above algorithm, $y_n$ is the dual variable for $x_n$ at the $n$-th iteration,  $\partial v(x_n)$ and $\partial u^*(y_n)$ are defined by $\partial v(x_n) = \inf\{x_n^Ty - v^*(y)\}$ and $\partial u^*(y_n)= \inf\{x^Ty_n - u(x): y \in \mathbb{R}^{2K} \}$ where $v^*(y)$ denotes the conjugate function of $v(x)$ which is defined by $v^*(y)=\inf \{ x^Ty-v(x) : x \in \mathbb{R}^{2K} \}$. The algorithm iterates until a convergence criterion is met (i.e., the change of the optimal $x$ is smaller than $\epsilon$).

\section{Simulation Results and Discussions}
\begin{table}[ht]
    \centering
    \caption{Simulation parameters.}
    \begin{tabular}{l l}
\hline\hline
     \textbf{Parameter} & \textbf{Value}\\\hline\hline
     Room Dimension     & 5 m $\times$ 5 m $\times$ 3 m  \\\hline
      LED power, $P_{\text{LED}}$ & 1W \\ \hline
      LED semi-angle at half illuminance, $\phi_{1/2}$    & $60^\circ$ \\\hline
      Energy conversion efficiency, $\eta$     & 0.44  \\\hline
      PD active area, $A_R$     & 1 $\text{cm}^2$  \\\hline
      PD responsivity, $R_P$     & 0.54 A/W  \\\hline
      PD field of view (FoV), $\Psi$     & $60^\circ $  \\\hline
     Optical filter gain, $T_s(\varphi_k)$      & 1  \\\hline
     Refractive index of concentrator, $\kappa$ &1.5  \\\hline
     Noise power, $\sigma_{k, \text{dl}}^2$, $\sigma_{k, \text{ul}}^2$ , $\sigma_{\text{E}, \text{dl}}^2$, $\sigma_{\text{E}, \text{ul}}^2$     & $10^{-14}$ W \\\hline
    \end{tabular}
    \label{tab: System parameters}
\end{table}

\begin{figure}[ht]
    \centering
    \includegraphics[width=0.45\textwidth]{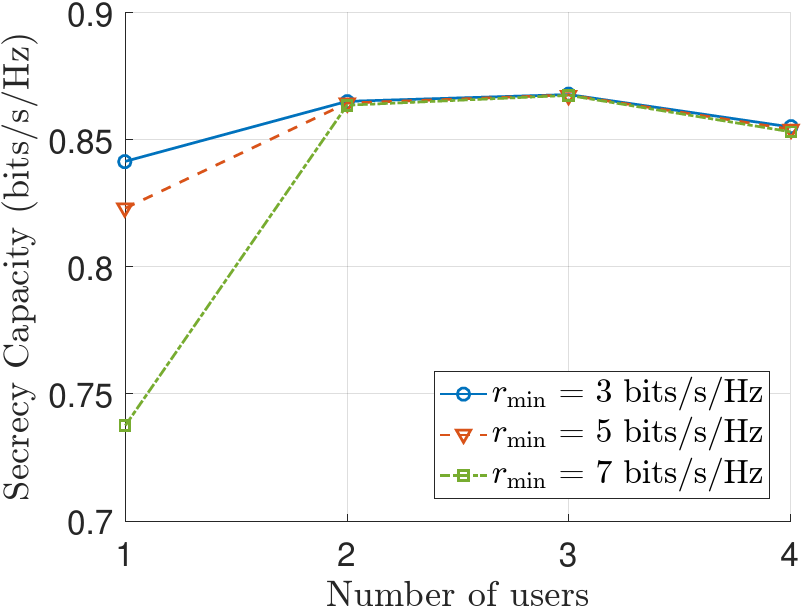}
    \caption{Secrecy capacity for different numbers of users}
    \label{fig:3}
\end{figure}

This section presents numerical results that illustrate the theoretical analyses of the optimized secrecy capacity in the UL.  The simulation parameters are given in Table \ref{tab: System parameters}. Figure \ref{fig:3} depicts the trade-off between the secrecy capacity of the UL RF link and the target sum rate of the DL VLC link for different numbers of users. For a single-user scenario, as the sum rate in the VLC link increases, the secrecy capacity in the RF link decreases. This occurs because spending more time transmitting data in the VLC link reduces the time available for energy harvesting, resulting in lower harvested power and consequently a reduced secrecy capacity in the RF link. However, when the number of users increases, changing the target VLC sum rate does not lead to a noticeable change in the secrecy capacity. 

\begin{figure}[ht]
    \centering
    \includegraphics[width=0.45\textwidth]{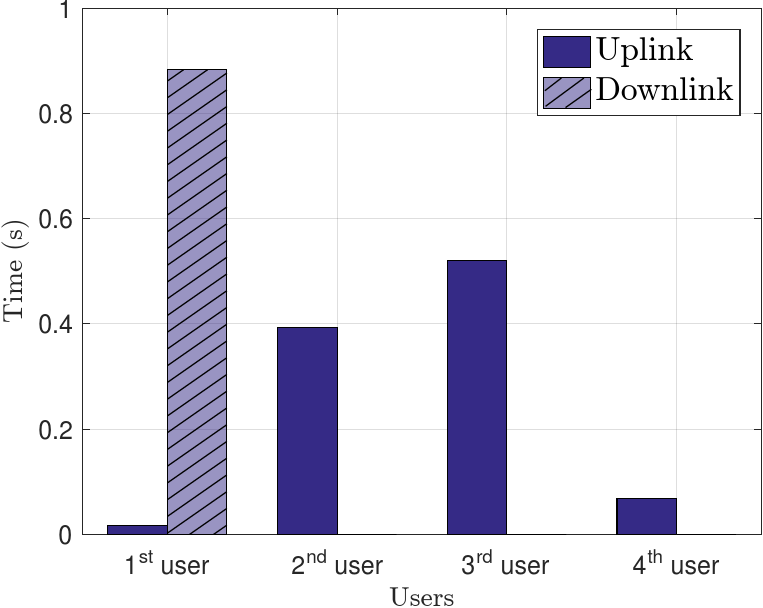}
    \caption{Optimal DL and UL time slots for 4 users.}
    \label{fig:4}
\end{figure}

Figure \ref{fig:4} illustrates the optimized time slot intervals for UL and DL links when the number of users is 4. The UL time slot is influenced by the channel gain of the DL, i.e., better channel gain of the VLC link allows for more transmission time in the RF link. For the downlink time frame, the transmission time is primarily allocated to the user with the worst channel gain. According to (\ref{eq:12}), since the secrecy capacity is inversely proportional to the DL time, the secrecy capacity reaches its maximum when the DL time is near zero. In the case of 4 users, to satisfy the constraint in (\ref{eq:11}), only one user has a significant DL time (i.e., the $1^{\text{st}}$ user) while the others have DL times close to zero to maximize the secrecy capacity. This leads to an unfairness among users' DL rates. 
\section{Conclusion}
This paper studied the secrecy capacity of the UL RF transmission in a hybrid RF-VLC system with SLIPT given a predefined target for the sum rate of the DL VLC channel. To maximize the secrecy performance, an optimization problem was formulated to solve the optimal DL and UL time slots for all users. Future work will address the issue of unfairness among users' DL rates.
\bibliographystyle{unsrt}
\bibliography{references}

\end{document}